\documentclass[11pt,twoside]{article}
\newcommand{\doi}[1]{doi: #1}
\newcommand{\pasp}{PASP}
\newcommand{\apj}{ApJ}
\usepackage{asp2021}

\aspSuppressVolSlug
\resetcounters

\begin{document}

\title{A New Plotly-Dash-based Query Infrastructure for the Keck Observatory Archive}

\author{R.~Moseley,$^1$ G.~Bruce~Berriman,$^1$ Christopher~R.~Gelino,$^1$ 
 John~C.~Good,$^1$ Meca~Lynn,$^1$ Melanie~Swain,$^1$ and Toba~Oluyide$^1$}
\affil{$^1$Caltech/IPAC-NExScI,~Pasadena,~CA~91125,~USA; \\
\email{moseley@ipac.caltech.edu}}

\paperauthor{R.~Moseley}{moseley@ipac.caltech.edu}{0000-0002-0759-0475}{Caltech}{IPAC/NExScI}{Pasadena}{CA}{91125}{USA}
\paperauthor{G.~Bruce~Berriman}{gbb@ipac.caltech.edu}{0000-0001-8388-534X}{Caltech}{IPAC/NExScI}{Pasadena}{CA}{91125}{USA}
\paperauthor{C.~R.~Gelino}{cgelino@ipac.caltech.edu}{0000-0001-5072-4574}{Caltech}{IPAC/NExScI}{Pasadena}{CA}{91125}{USA}
\paperauthor{J.~C.~Good}{jcg@ipac.caltech.edu}{0009-0003-3906-719X}{Caltech}{IPAC/NExScI}{Pasadena}{CA}{91125}{USA}
\paperauthor{M.~Lynn}{mlynn@ipac.caltech.edu}{0009-0008-7417-3170}{Caltech}{IPAC/NExScI}{Pasadena}{CA}{91125}{USA}
\paperauthor{M.~Swain}{mswain@ipac.caltech.edu}{0000-0003-4557-1192}{Caltech}{IPAC/NExScI}{Pasadena}{CA}{91125}{USA}
\paperauthor{T.~Oluyide}{toluyide@ipac.caltech.edu}{}{Caltech}{IPAC/NExScI}{Pasadena}{CA}{91125}{USA}



\begin{abstract}
The Keck Observatory Archive (KOA) curates all observational data acquired at the W. M. Keck Observatory. The archive is expected to grow rapidly as complex new instruments are commissioned and as the expectations of archive users have expanded. In response, KOA has implemented a new Python-based, VO-compliant query infrastructure. This work is a continuation of the architectural design and technology selection identified at ADASS 2024. We have deployed real-time ingestion of newly acquired data and a dedicated interface for observers to manage these data. Our ADASS 2024 poster identified the new technologies chosen: Plotly-Dash, a low-code framework that exploits event-driven callbacks to simplify the handling of user interactions; R-tree spatial indexing to speed up spatial searches by $\times 20$; a VO-compliant TAP middleware, already in use at the NASA Exoplanet Archive and NEID archive; and \texttt{mViewer}, a visualization engine in the Montage Image Mosaic toolkit that is optimized for astronomy images.

These technologies will underpin new services that can be hosted on web pages or in Jupyter notebooks, and when completed, will replace the current query infrastructure. We have completed two new services now in beta release. The first is the Data Discovery Service, a web-based dashboard that returns spatial and temporal queries of the entire archive in seconds. It supports filtering observations by keywords, previewing results in an interactive data grid, and visualizing images, and it offers data downloads. The second is a Jupyter notebook that performs interactive visualization of Keck observations of protostars in the Rho Oph Dark Cloud and uses data from CDS and IRSA, as well as KOA.

\end{abstract}



\section{Introduction}
The Keck Observatory Archive (KOA) manages over 30 years of raw and reduced data, from all current and decommissioned W. M. Keck Observatory instruments. The archive is currently modernizing its operations and query infrastructure to respond to changing directions in research and to serve massive datasets from the next generation of instruments. Modernization prioritizes widely adopted, open-source, and sustainable Python technologies wherever possible.
This paper describes two services that use the core technologies, described in \citet{Moseley} and summarized below, which have been demonstrated in the following two YouTube videos:

\begin{itemize}
    \item A Data Discovery Service (DDS) that queries the archive in seconds\footnote{\url{https://www.youtube.com/watch?v=RMtqsMjeh-Y}}
    \item A Jupyter notebook that uses the DDS to discover Keck data and integrate them with remote datasets, accessed through Virtual Observatory (VO) protocols\footnote{\url{https://www.youtube.com/watch?v=UC6-tuWWYQo}}.
\end{itemize}

\section{Data Discovery Service}

The Data Discovery Service (DDS) is a major part of the modernization of the KOA query infrastructure. The DDS takes a "low-code," Python-centric approach to architecture through the Plotly-Dash framework. This framework uses event-driven callbacks to manage complex client-server interactions, including coordinate resolution and quick spatial searches, without the overhead of maintaining a decoupled JavaScript front end. Figure \ref{ex_fig1} shows a block diagram of the DDS functionality.

The video demonstration performs a spatial search of a 6\deg radius centered on M51, which has been observed with every instrument at the Keck Observatory since it began operating in 1994. The DDS queried the entire archive of six million records in approximately 15 seconds. The return is shown in Figure \ref{ex_fig2}. The left column shows an instrument-by-instrument summary; on the right is an interactive tabbed display of the metadata.

\articlefigure[width=0.5\textwidth]{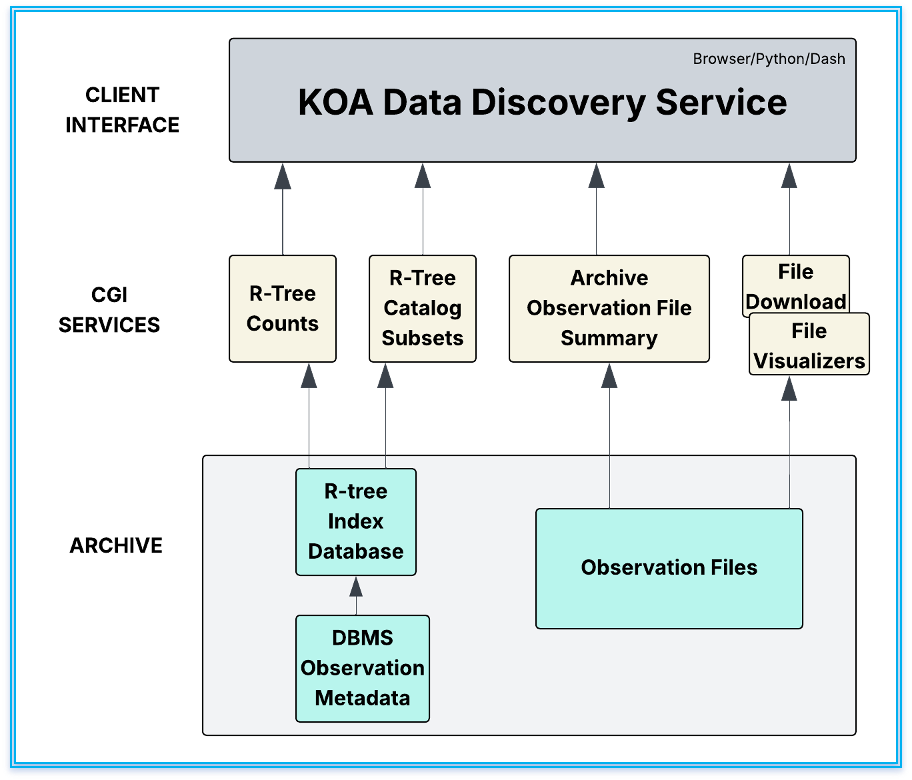}{ex_fig1}{Overview of the Data Discovery Service (DDS) functional design}

\articlefiguretwo{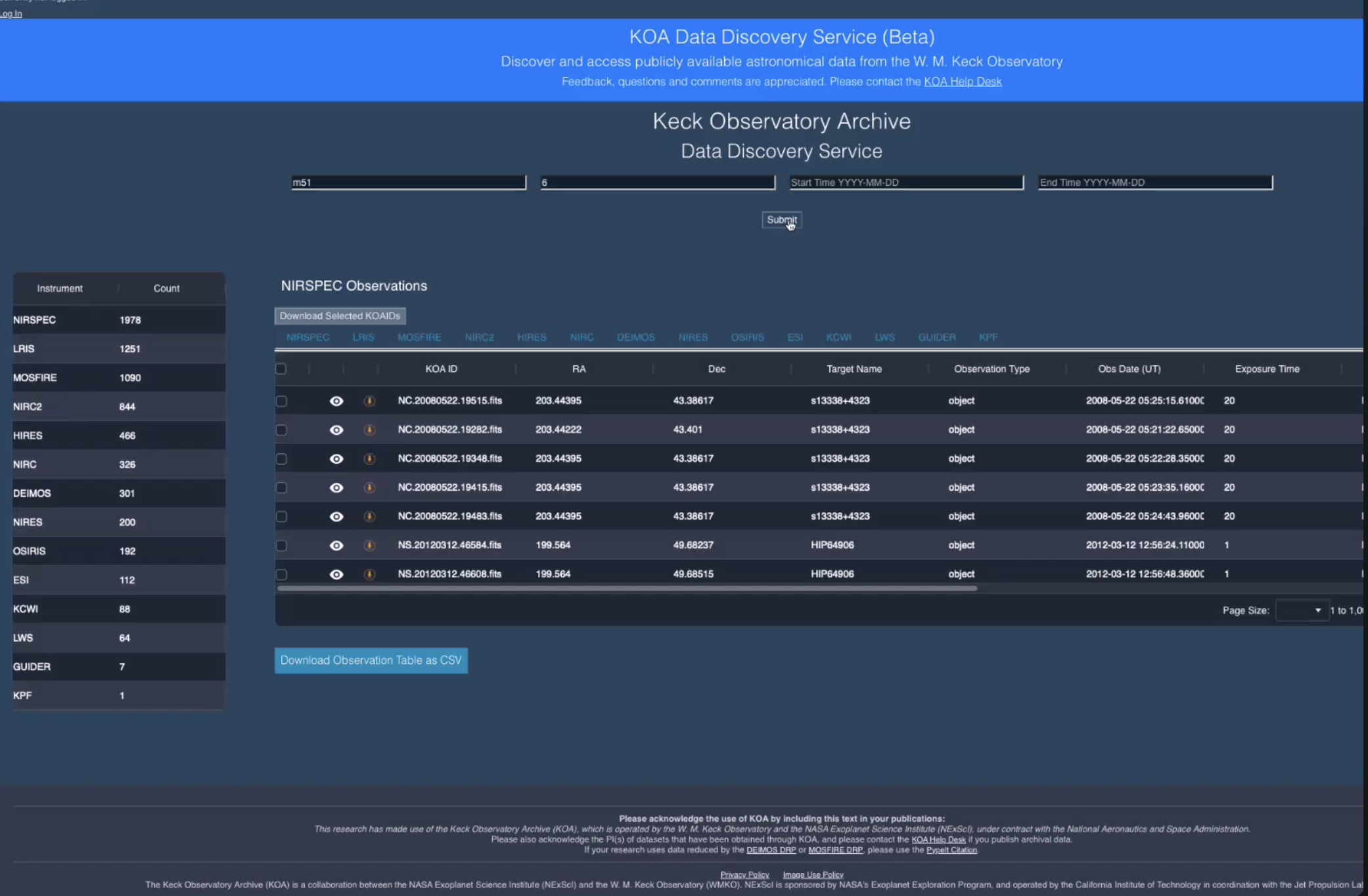}{F1_f2_b}{ex_fig2}{The Data Discovery Service (DDS) return page. \emph{Left:} The main search and results dashboard showing the instrument-based observation summary and filtered data table \emph{Right:} The integrated \texttt{mViewer} component displaying a DEIMOS FITS image preview with interactive histogram and stretch controls}

The DDS uses the following technologies:

\begin{itemize}
    \item \texttt{nexsciTAP}, a Python Table Access Protocol (TAP) implementation that provides a VO-compliant mechanism for handling Astronomy Data Query Language (ADQL) requests.
    \item R-tree spatio-temporal indices that approximate complex instrument geometries using minimum bounding rectangles (MBRs). Grouping nearby objects into hierarchical MBRs \citep{10.1145/602259.602266} instantiated as memory-mapped files\footnote{\url{https://github.com/Caltech-IPAC/Montage}}. R-tree indices allow fast searches, $\times 20$ faster than current services.
    \item A Dash React component wraps the \texttt{mViewer} module of the Montage Image Mosaic Engine \citep{2017PASP..129e8006B} to provide an adaptive, histogram-based visualizer.
    \item \texttt{wget} shell scripts for bulk data retrieval.

\end{itemize}

\section{KOA DDS Notebook}
We have delivered an example Jupyter notebook that uses the DDS to interact with remote observations of Rho Ophiuchi (Rho Oph). Two screenshots from the video\footnote{\url{https://www.youtube.com/watch?v=UC6-tuWWYQo}} are shown in Figure \ref{ex_fig3}. This second demonstration details the following workflow: after resolving the coordinates of Rho Oph with SIMBAD and defining a search radius, the DDS discovers all Keck observations in Rho Oph, retrieves a catalog of protostars in Rho Oph from VizieR \citep{2017ApJ...835....3L}, and overlays these data on a Spitzer IRAC mosaic acquired from IRSA 

\articlefiguretwo{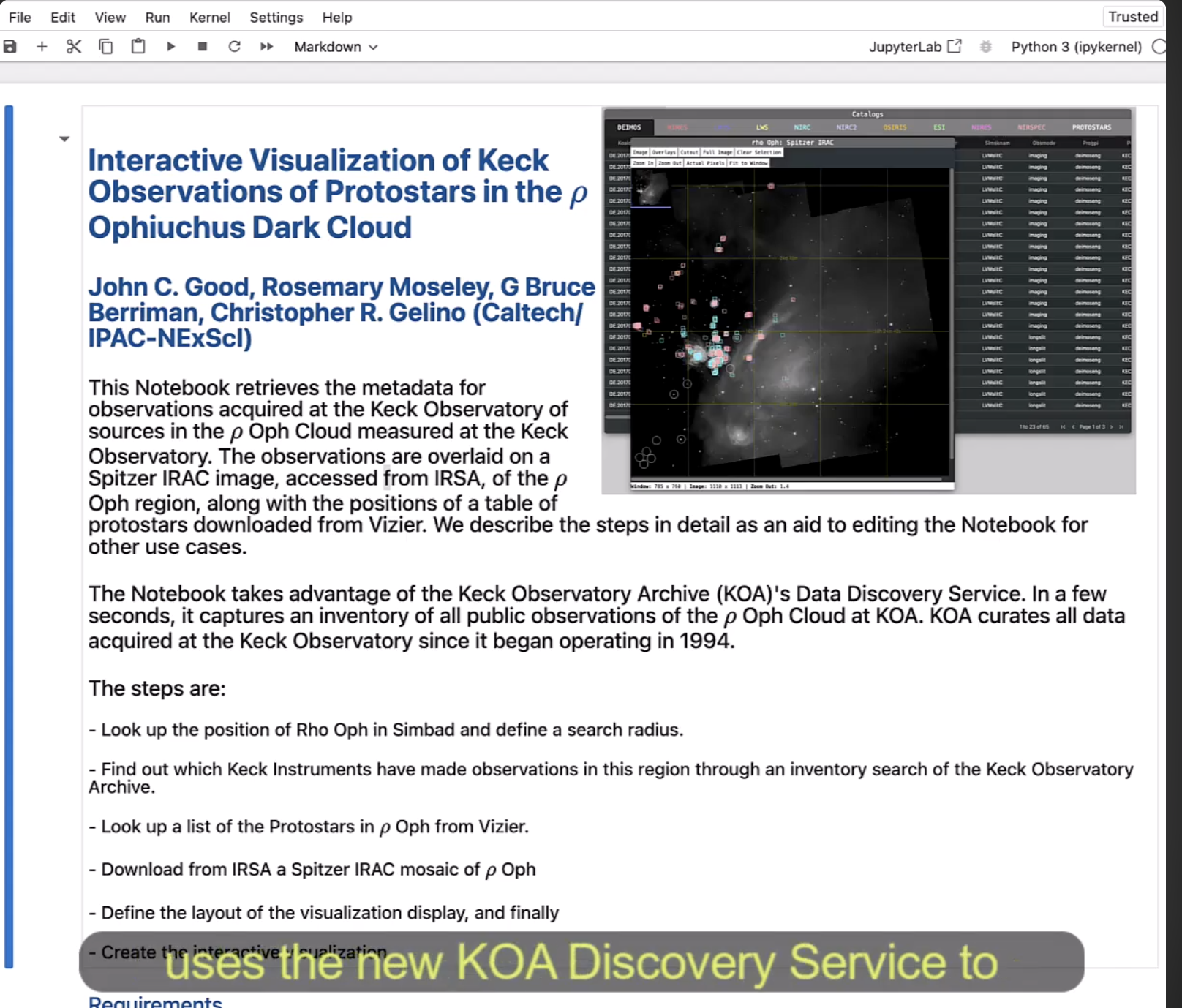}{F1_f3_b}{ex_fig3}{Jupyter Notebook integration with the DDS API. \emph{Left:} Notebook usage of Keck observations and external VO catalogs within a Jupyter environment \emph{Right:} The resulting interactive visualization overlaid on the Spitzer IRAC mosaic}

\section{Conclusion}
The Plotly-Dash framework provides a powerful mechanism for building a new query infrastructure for KOA. The DDS and the associated notebook are examples of powerful new services that can be built with the new query infrastructure, and will deliver new tools to serve the needs of the rapidly expanding archive. 

\acknowledgements The Keck Observatory Archive (KOA) is a collaboration between the NASA Exoplanet Science Institute (NExScI) and the W. M. Keck Observatory (WMKO). NExScI is sponsored by the NASA Exoplanet Exploration Program and operated by the California Institute of Technology in coordination with the Jet Propulsion Laboratory (JPL).

The observatory was made possible by the generous financial support of the W. M. Keck Foundation. The authors wish to recognize and acknowledge the very significant cultural role and reverence that the summit of Mauna Kea has always had within the indigenous Hawaiian community. We are most fortunate to have the opportunity to conduct observations from this mountain.


\begin{thebibliography}{}

\bibitem[Berriman \& Good(2017)]{2017PASP..129e8006B}
{Berriman}, G.~B., \& {Good}, J.~C. 2017, \pasp, 129, 058006. \doi{10.1088/1538-3873/aa5456}

\bibitem[Guttman(1984)]{10.1145/602259.602266}
Guttman, A. 1984, in Proceedings of the 1984 ACM SIGMOD International Conference on Management of Data, SIGMOD '84 (New York, NY, USA: ACM), 47--57. \doi{10.1145/602259.602266}

\bibitem[Lindberg et~al.(2017)]{2017ApJ...835....3L}
{Lindberg}, J.~E., {Charnley}, S.~B., {J{\o}rgensen}, J.~K., {Cordiner}, M.~A., \& {Bjerkeli}, P. 2017, \apj, 835, 3. \doi{10.3847/1538-4357/835/1/3}

\bibitem[Moseley et~al.(2024)]{Moseley}
{Moseley}, R., {Berriman}, G.~B., {Gelino}, C.~R., {Good}, J.~C., \& {Oluyide}, T. 2024, in ADASS XXXIV, vol. TBD of ASP Conf. Ser., 999. \url{https://arxiv.org/abs/2412.12356}

\end{thebibliography}


\end{document}